\documentclass[twocolumn,showpacs,preprintnumbers]{revtex4}
\usepackage{amssymb}
\usepackage{graphicx}
\usepackage{amsmath}
\usepackage{hyperref}
\usepackage{epsfig,amscd,amssymb}

\begin{document}

\title{On the Kinetic Equation and Electrical Resistivity in Systems with Strong Spin- Hole Interaction}

\author{A.F.Barabanov$^{\ast }$}
\affiliation{Institute for High Pressure Physics, Russian Academy
of Sciences, Troitsk 142190, Moscow Region, Russia}
\email{abarab@bk.ru}
\author{A.M. Belemuk}
\affiliation{Institute for High Pressure Physics, Russian Academy
of Sciences, Troitsk 142190, Moscow Region, Russia}
\author{L.A. Maksimov}
\affiliation{Russian Research Centre Kurchatov Institute, pl.
Kurchatova 1, Moscow; 123182, Russia}

\begin{abstract}
The problem of constructing the kinetic equation with the
description of motion of a hole in systems with strong spin- hole
interaction (such as high- temperature superconductors) in terms
of the spin polaron has been considered in the framework of the
regular antiferromagnetic $s-d$ model. It has been shown by the
example of the electrical resistivity that kinetics is determined
by the properties of the bands of the spin polaron (rather than
"bar hole") and their quasiparticle residues $Z_{k}$. The cases of
low and optimal doping of the $CuO_{2}$ plane have been
considered. It has been shown that the rearrangement of the
spectrum of the lower polaron band, as well as the strong doping
dependence of the quasiparticle residues $Z_{k}$ is decisive in
the unified consideration of these cases.
\end{abstract}


\maketitle

It is known that the normal state of high- temperature
superconductors is characterized by a complex behavior of the
spectral and transport properties due to the strong interaction of
the carriers with the spin subsystem \cite{LeeRMP06,
DamascelliRMP03,Sachdev04+}. This refers to the nontrivial
evolution of the hole and spin subsystems with increasing doping,
when the system passes from the Mott dielectric to the metallic
state \cite{Ando04}.

The overwhelming majority of the studies \cite{Stojkovic97} on the
microscopic description of the kinetics in high- temperature
superconductors was devoted to the case of the optimal doping and
was based on the concept of the almost antiferromagnetic Fermi
liquid described by the spin- fermion Hamiltonian $\hat{H}$ on the
square lattice
\begin{gather}
\hat{H}_{t}= \hat{H}+\hat{H}_{f}, \quad \hat{H}_{f}=
-\widehat{d}^{x}E^{x}, \nonumber \\
\hat{H}=\hat{H}_{00}+ \widehat{J},\quad \hat{H}_{00}=
\hat{H}_{0h}+ \hat{I}, \label{1}
\end{gather}
\begin{gather}
\hat{H}_{0h}= \sum_{\mathbf{k},\sigma } \varepsilon_{\mathbf{k}}
a_{\mathbf{k}\sigma }^{\dagger }a_{\mathbf{k}\sigma }, \nonumber \\
\hat{I}= \frac{1}{2} I(1-p) \sum \limits_{\mathbf{R},\mathbf{g}}
S_{\mathbf{R+g}}^{\alpha }S_{\mathbf{R}}^{\alpha}+ \frac{1}{2}Ip \sum
\limits_{\mathbf{R},\mathbf{d}} S_{\mathbf{R+d}}^{\alpha }S_{\mathbf{R}}^{\alpha },  \label{2} \\
\widehat{J}= \frac{J}{\sqrt{N}} \sum
\limits_{\mathbf{k},\mathbf{q},\gamma _{1},\gamma
_{2}}a_{\mathbf{k}+\mathbf{q},\gamma _{1}}^{\dagger
}S_{\mathbf{q}}^{\alpha }\hat{\sigma}_{\gamma _{1}\gamma
_{2}}^{\alpha }a_{\mathbf{k}\gamma _{2}}=
\sum\limits_{\mathbf{q}}\Lambda _{\mathbf{q}}^{\alpha
}S_{\mathbf{q}}^{\alpha }, \nonumber \\
\Lambda _{\mathbf{q}}^{\alpha
}=\frac{J}{\sqrt{N}}\sum\limits_{\mathbf{k},\gamma _{1},\gamma
_{2}}a_{\mathbf{k}+\mathbf{q},\gamma _{1}}^{\dagger
}\hat{\sigma}_{\gamma _{1}\gamma _{2}}^{\alpha
}a_{\mathbf{k}\gamma _{2}}  \label{3}
\end{gather}

The term $\hat{H}_{0h}$ in the Hamiltonian $\hat{H}_{00}$
describes the bar Fermi carriers and contains the spectrum
$\varepsilon _{\mathbf{k}}$ of bar holes; $\hat{I}$ corresponds to
the frustrated antiferromagnetic interaction between $S=1/2$
spins, $\mathbf{g}$ and $\mathbf{d}$ are the vectors of the first
and second neighbors, respectively; and $I_{1}=(1-p)$ and
$I_{2}=pI$, where $p$ ($0\leq p\leq 1$) is the frustration
parameter, are the respective antiferromagnetic exchange
constants. The term $\widehat{J}$ describes the of the the
carriers with the subsystem of localized spins
$\mathbf{S}_{\mathbf{R}}$ ($\widehat{\sigma }^{\alpha }$ are the
Pauli matrices and the summation over repeated Cartesian
superscripts $\alpha $ and spin subscripts $\gamma _{1}$ and
$\gamma _{2}$ is implied). The total Hamiltonian $\hat{H}_{t}$
includes the interaction $\hat{H}_{f}$ with the electric field
$E$, and $\widehat{d}^{x}$ is the dipole momentum operator.

In order to adequately describe the temperature dependence of the
electrical resistivity $\rho (T)$ as well as the Hall coefficient
$R_{H}(T)$) by solving the kinetic equation in the almost
antiferromagnetic liquid model, the spectrum $\varepsilon
_{\mathbf{k}}$ in Hamiltonian $\hat{H}_{0h}$ is always changed to
the spectrum $E_{\mathbf{k}}^{(1)}$ of the lowest quasiparticle
band of the spin polaron, and the operators $a_{\mathbf{k}\sigma}$
in $\hat{H}_{0h}$ are remained fermion operators. The spectrum
$E_{\mathbf{k}}^{(1)}$ corresponds to a large Fermi surface and is
well measured in experiments on angle resolved photoemission
spectroscopy (ARPES). Such a change is not obvious and inevitably
leads to the incorrect number of holes $\widetilde{n}_{h}\approx
1.2$ instead of the real relation $n_{h}\lesssim 0.2$. However,
for the physical quantities appearing in the kinetic equation and
in the expression for the average current, this change seems
reasonable, because the kinetic must be determined by the
velocities of quasiparticles,
$\mathbf{V}_{\mathbf{k}}^{(1)}=\partial
E_{\mathbf{k}}^{(1)}/\partial \mathbf{k}$, of the lower polaron
band rather than by the velocities of bar holes,
$\mathbf{v_{k}}=\partial \varepsilon _{\mathbf{k}}/\partial
\mathbf{k}$.

In this study, we present the derivation of the kinetic equation
that is based on the Hamiltonian (\ref{1}) and spin- polaron
concept and leads to the observed $\rho (T)$ dependence for the
large Fermi surface and $n_{h}\lesssim 0.2$ (due to the small
quasiparticle residues $Z_{\mathbf{k}}^{(1)}\lesssim 0.2$ of the
lower polaron band in the Green's function of the bar hole), as
well as to the natural appearance of the quasiparticle velocities
$\mathbf{V}_{\mathbf{k}}^{(1)}=\partial
E_{\mathbf{k}}^{(1)}/\partial \mathbf{k}$.

The carriers are described in the multipole approximation, i.e.,
in the sufficiently complete basis of the spin polaron. Under the
assumption that doping stimulates the frustration of the spin
subsystem, the cases of small and optimal doping of the $CuO_{2}$
plane are discussed. The simultaneous consideration of these cases
is possible only with allowance for the strong rearrangement of
the residue function $Z_{\mathbf{k}}^{(1)}$ with doping.

Let us first consider the equilibrium Green's functions for bar
holes and introduce spin polaron states by an example of the
simplest approximation.

At characteristic values $J\simeq 0.1$ eV, the Hamiltonian
$\hat{J}$ corresponds to the strong interaction of bar hole with
the spin subsystem. As a result, elementary charge excitations
must be described by the spin polaron, which is represented as a
superposition of the bar- hole operator $a_{\mathbf{k}\sigma }$
and the spin polaron operators describing "dressing" of the
$a_{\mathbf{k}\sigma }$ to the operators of the spin subsystem.
The problem is solved with the use of the Mori- Zwanzig projection
method for Green's functions. The method implies the choice of a
finite set of the basis operators, which must include the pairing
of the bar hole with localized spins from the very beginning.

It is known \cite{Bar2001} that the minimum "good" site set is the
following set of the basis operators:
\begin{gather}
\varphi _{\mathbf{r} \sigma}^{(1)}=a_{\mathbf{r} \sigma}, \quad
\varphi _{\mathbf{r} \sigma}^{(2)}= S^{\alpha}_{\mathbf{r}} \hat
\sigma^{\alpha}_{\sigma \sigma_1}{\,}a_{\mathbf{r} \sigma_1},
\label{fi12} \\
\varphi _{\mathbf{r}\sigma
}^{(3)}=\frac{1}{N}\sum\limits_{\mathbf{\rho,q\in \Omega
}}{\,}e^{i\mathbf{q\rho }}{\,}S_{\mathbf{r+\rho }}^{\alpha
}\hat{\sigma}_{\sigma \sigma _{1}}^{\alpha
}{\,}a_{\mathbf{r}\sigma _{1}}, \nonumber \\
\varphi _{\mathbf{r}\sigma
}^{(4)}=\frac{1}{N}\sum\limits_{\mathbf{\rho ,q\in \Omega
}}{\,}e^{i\mathbf{q\rho }}{\,}S_{\mathbf{r+\rho }}^{\alpha
}\hat{\sigma}_{\sigma \sigma _{1}}^{\alpha
}{\,}S_{\mathbf{r}}^{\beta }\hat{\sigma}_{\sigma _{1}\sigma
_{2}}^{\beta }{\,}a_{\mathbf{r}\sigma _{2}},  \label{fi34} \\
\Omega =\{\mathbf{q}:\quad |\pm (\pi /g)-q_{x,y}|<L\}. \nonumber
\end{gather}

The first two operators $\varphi _{\mathbf{r}\sigma }^{(1)}$ and
$\varphi _{\mathbf{r}\sigma }^{(2)}$ can be treated as local spin-
polaron operators, the following two operators
$\varphi_{\mathbf{r}\sigma }^{(3)}$ and
$\varphi_{\mathbf{r}\sigma}^{(4)}$ correspond to the spin polaron
of the intermediate radius and describe the pairing of local
polaron operators $\varphi _{\mathbf{r}\sigma }^{(1)}$ and
$\varphi _{\mathbf{r}\sigma }^{(2)}$ with the spin wave operators
\begin{equation*}
S_{\mathbf{q}}^{\alpha
}=\frac{1}{\sqrt{N}}\sum\limits_{\mathbf{\rho
}}e^{i\mathbf{q(r+\rho )}}S_{\mathbf{r+\rho }}^{\alpha }.
\end{equation*}

A feature of operators (\ref{fi34}) is that they reflect the
pairing of spin waves $S_{\mathbf{q}}^{\alpha }$ with momenta
$\mathbf{q}$ close to the antiferromagnetic vector
$\mathbf{Q}=(\pi ,\pi)$, i.e., $\mathbf{q}$ values fill region
$\mathbf{\Omega}$ consisting of four $L\times L$ squares in the
corners of the first Brillouin zone (in what follow, we set
$\Omega =L\times L=0.25(\pi /g)^{2}$). Pairing with such spin
waves $S_{\mathbf{q}}^{\alpha }$ takes into account the sharp peak
of the spin- spin structure factor in the region close to the
antiferromagnetic vector $\mathbf{Q}$ and is responsible for the
splitting of the lower quasiparticle band appearing in the local
polaron approximation \cite{Bar2001}. Moreover, the inclusion of
the finite region $\mathbf{\Omega }$ is necessary for describing
the correct passage to the limit $T\rightarrow 0$.

The standard projection procedure for solving the equations for
Green's functions in the momentum representation for operators
(\ref{fi12}) and (\ref{fi34}) provides four bands of the spin
polaron $E_{\mathbf{k}}^{(s)}$ ($s$ is the band number), an
explicit expression for the Green's function of bar hole,
$G_{h}(\mathbf{k},\omega )=\ll a_{\mathbf{k}\sigma
}|a_{\mathbf{k}\sigma }^{\dagger }\gg _{\omega }$, the expression
for the number of bar holes, $n_{\mathbf{k}\sigma }$ in terms of
quasiparticle residues $Z_{\mathbf{k}}^{(l)}$, and makes it
possible to represent Hamiltonian $\hat{H}$ given by Eq.(\ref{1})
in the polaron basis as $\hat{H}_{p}$:
\begin{gather}
G_{h}(\mathbf{k},\omega
)={\,}\sum_{s=1}^{4}\frac{Z_{\mathbf{k}}^{(s)}}{\omega
-E_{\mathbf{k}}^{(s)}}, \nonumber \\
n_{\mathbf{k}\sigma}= \langle a_{\mathbf{k}\sigma }^{\dagger
}a_{\mathbf{k}\sigma} \rangle=
\sum_{s=1}^{4}Z_{\mathbf{k}}^{(s)}n_{F}(E_{\mathbf{k}}^{(s)}); \label{a5} \\
n_{h}=\sum_{\mathbf{k,}\sigma,s}Z_{\mathbf{k}}^{(s)}n_{F}(E_{\mathbf{k}}^{(s)}).
\nonumber
\end{gather}

Here $n_F(E_{\mathbf{k}})=(e^{(E_{\mathbf{k}}-\mu)/T}+ 1)^{-1}$,
where  $\mu$ is the chemical potential,
\begin{gather}
\varphi _{\mathbf{k},\sigma }^{(i)}= U_{ij}^{-1}(\mathbf{k})\alpha
_{\mathbf{k}\sigma }^{(j)}, \quad \langle \langle \alpha
_{\mathbf{k}\sigma }^{(s)}|\alpha _{\mathbf{k}\sigma }^{(s^{\prime
})\dagger }\rangle \rangle _{\omega }= \frac{1}{\omega
-E_{\mathbf{k}}^{(s)}}\delta _{ss^{\prime }}, \nonumber \\
Z_{\mathbf{k}}^{(s)}=\overline{U_{s1}^{-1}(\mathbf{k})}U_{s1}^{-1}(\mathbf{k}),
\quad u_{\mathbf{k}s}=U_{s1}^{-1}(\mathbf{k}), \quad
Z_{\mathbf{k}}^{(s)}=u_{\mathbf{k}s}^{2}, \label{a7} \\
\sum\limits_{s}Z_{\mathbf{k}}^{(s)}=1, \quad a_{\mathbf{k},\sigma
}=\sum_{s=1}^{4}u_{\mathbf{k}s}\alpha _{\mathbf{k}\sigma }^{(s)},
\nonumber \\
\hat{H}_{p}= \sum_{\mathbf{k},\sigma ,s}E_{\mathbf{k}}^{(s)}\alpha
_{\mathbf{k}\sigma }^{(s)\dagger }\alpha _{\mathbf{k}\sigma
}^{(s)}=\hat{H}_{00}+\widehat{J}_{p}, \nonumber \\
\widehat{J}_{p}=\widehat{P}\widehat{J}\widehat{P}, \quad
\widehat{P}=\sum_{\mathbf{k,}s,\gamma }|\alpha _{\mathbf{k,}\gamma
}^{(s)}\rangle \langle \alpha _{\mathbf{k,}\gamma }^{\dagger (s)}|
\label{a8}
\end{gather}

Here, $\widehat{P}$ is the projection operator on the polaron
space. The matrix $U_{ij}^{-1}(\mathbf{k})$ is expressed in the
explicit form in terms of the spin correlation functions, which
are in turn determined in terms of the susceptibility
$\chi(\mathbf{q},\omega)$ of the frustrated antiferromagnetic spin
subsystem.

The Hamiltonian $\hat{H}_{p}$ contains the operator
$\widehat{J}_{p}$, which is a component of $\widehat{J}$ that is
responsible for the formation of polaron from a bar hole. Operator
$\widehat{J}_{p}$ includes only those matrix elements of
$\widehat{J}$  which describe the polaron scattering processes
without change in the quasimomentum, i.e., $\widehat{J}_{p}\alpha
_{\mathbf{k}\sigma }^{(s1)}\ =>$ $\alpha _{\mathbf{k}\sigma
}^{(s2)}$ processes. This scattering processes are taken into
account in the projection- approach description of the formation
of polarons.

Comparison of the projection- method results for the a complex
spin polaron \cite{Bar2001} with the self- consistent Born
approximation (SCBA) calculations (at $T=0$) \cite{Kuzan}
certainly indicates that the lower band $E_{\mathbf{k}}^{(1)}$ and
residues $Z_{\mathbf{k}}^{(1)}$ well reproduce the SCBA peak and
its intensity. The upper three bands $E_{\mathbf{k}}^{(2)}$,
$E_{\mathbf{k}}^{(3)}$, and $E_{\mathbf{k}}^{(4)}$ effectively
describe the incoherent part of $A_{incoh}(\mathbf{k},\omega)$ of
the total hole SCBA spectral function
\begin{equation*}
A_{SCBA}(\mathbf{k},\omega
)=Z_{\mathbf{k}}^{(1)}{\,}\delta (\omega
-E_{\mathbf{k}}^{(1)})+A_{incoh}(\mathbf{k},\omega )
\end{equation*}

The motion of the bar hole under the action of, e.g., the external
electric field in terms of polaron operators is the motion of spin
polaron $\alpha _{\mathbf{k}\sigma }^{(s)}$ simultaneously in four
bands with the velocity $\mathbf{V}_{\mathbf{k}}^{(s)}=\partial
E_{\mathbf{k}}^{(s)}/\partial \mathbf{k}$.

The operator $\widehat{\widetilde{J}}$ such that $\hat{H}=
\hat{H}_{p}+\widehat{\widetilde{J}}$ should be introduced in the
kinetic equation. Here, $\widehat{\widetilde{J}}$ should be
treated as an operator including the matrix elements of
$\widehat{J}$, which lead to polaron- polaron scattering
$\widehat{\widetilde J} \alpha_{\mathbf{k}\sigma }^{(s1)}$ $=>$ $\alpha
_{\mathbf{k+q},\sigma }^{(s2)} S_{\mathbf{-q}}^{\alpha }$ with
change in the quasimomentum $\mathbf{q\neq 0}$ and with the
simultaneous excitation of the spin subsystem. This vertex
obviously must describe the collision term in the kinetic equation
for polarons in the second order in $J$.

The Hamiltonian in the polaron representation takes the form (from
now on, we take term $\hat I$ out of the polaron Hamiltonian $H_p$
in explicit way)
\begin{equation} \label{4}
\hat{H}= \hat{H}_{0}+\widehat{\widetilde{J}}, \quad
\hat{H}_{0}=\hat{H}_{00}+ \widehat{J}_{p}= \hat{H}_{p}+\hat{I}
\end{equation}
Here,
\begin{equation*}
\hat{H}_{p}=\sum_{\mathbf{k},\sigma ,s}E_{\mathbf{k}}^{(s)}\alpha
_{\mathbf{k}\sigma }^{(s)\dagger}\alpha _{\mathbf{k}\sigma}^{(s)},
\end{equation*}
where
\begin{gather}
\widehat{\widetilde{J}}= \sum\limits_{\mathbf{q\neq 0,}\alpha
}\Lambda _{\mathbf{q}}^{\alpha }S_{\mathbf{q}}^{\alpha },
\nonumber \\
\Lambda _{\mathbf{q}}^{\alpha
}=\frac{J}{\sqrt{N}}\sum\limits_{\mathbf{k},\gamma _{1},\gamma
_{2}}(\sum_{s1=1}^{4}u_{\mathbf{k+ q}s1}\alpha
_{\mathbf{k+q,}\gamma _{1}}^{\dagger (s1)})\hat{\sigma}_{\gamma
_{1}\gamma _{2}}^{\alpha }(\sum_{s2=1}^{4}u_{\mathbf{k}s2}\alpha
_{\mathbf{k}\gamma
_{2}}^{(s2)}), \label{5} \\
\hat{H}_{t}= \hat{H}+\hat{H}_{f}, \quad \hat{H}_{f}=
-\widehat{d}^{x}E, \nonumber \\
\widehat{d}^{x}=e\sum\limits_{\mathbf{k,k^{\prime }},\sigma
}x_{\mathbf{kk^{\prime }}}a_{\mathbf{k}\sigma }^{\dagger
}a_{\mathbf{k^{\prime }}\sigma },  \label{6}
\end{gather}
where
\begin{gather}
x_{\mathbf{kk^{\prime }}}= \langle
\mathbf{k}|\hat{x}|\mathbf{k^{\prime }}\rangle= (\Delta
_{\mathbf{k}^{^{\prime }}\mathbf{,k+q}}+\Delta
_{\mathbf{k}^{^{\prime }}\mathbf{,k-q}})/2q|_{\lim (q=>0)},
\nonumber \\
\mathbf{q}=(q,0); \quad x_{\mathbf{k,k}^{^{\prime
}}}=x_{\mathbf{k}^{^{\prime },}\mathbf{k}}.  \label{7}
\end{gather}

The transition to operators $\alpha _{\mathbf{k}\sigma }^{(s)}$ is
implied in the operator $\widehat{d}^{x}$ and the quasi-
inhomogeneous field $E$ directed along the $X$ axis is introduced
according to the Eq. (\ref{6}).

In order to obtain an expression for the electrical resistivity,
we use a variant of the linear- response theory in which the field
$E$ that ensures a fixed electrical current $j$ (rather than the
current at a fixed field) is sought. Deviation from equilibrium
($\kappa$ is the system deviation parameter specifying $j$) at the
initial time $t=0$ is characterized by a finite set of operators
$\hat{F}_{l}^{s}$ and the density matrix $\widehat{\widetilde{\rho
}}$ is specified in the form
\begin{gather}
\widehat{\widetilde{\rho }}= \hat{\rho}^{0}+ \widehat{\phi}^{0},
\quad \hat{\rho}^{0}=Z_{0}^{-1}\exp (-\hat{H}_{0}/T), \nonumber \\
\langle \widehat{A} \rangle= Sp\{\hat{\rho}^{0}\widehat{A}\}, \label{ro2} \\
\widehat{\phi }^{0}= \hat{\rho}^{0}\hat{F}, \quad \hat{F}=
\sum_{l,s}\eta _{l}^{s} \hat{F}_{l}^{s}, \nonumber \\
\langle \hat{F}_{l}^{s}\rangle _{0}= 0, \quad
\overline{\hat{F}_{l}^{s}}|_{t=0}= \sum_{l_{1},s_{1}}
\eta_{l_{1}}^{s_{1}} \langle \hat{F}_{l}^{s}
\hat{F}_{l_{1}}^{s_{1}} \rangle, \\
\hat{F}_{l}^{s}= \sum_{\mathbf{k},\sigma}
F_{l}^{s}(\mathbf{k)}\alpha _{\mathbf{k}\sigma
}^{(s)\dagger}\alpha _{\mathbf{k}\sigma }^{(s)}, \nonumber \\
[\hat{F}_{l}^{s}, H_{0}]= 0, \quad
[\hat{F}_{l}^{s},\hat{\rho}^{0}]=0.
\end{gather}

Note that the moments $F_{l}^{s}(\mathbf{k)}$ in the current state
are odd in $\mathbf{k}$ and the moment $\hat{F}_{1}^{1}=
\sum_{\mathbf{k},\sigma }\mathbf{V}_{\mathbf{k}}^{1}\alpha
_{\mathbf{k}\sigma }^{(1)\dagger }\alpha _{\mathbf{k}\sigma
}^{(1)}$ corresponds to the simplest one- moment approximation
associated with the lower polaron band.

The evolution equation for the density matrix $\hat{\rho}_{t}=
\widehat{\widetilde{\rho }}+\hat{\rho}_{t}^{\prime }$ has the form
$i\frac{\partial \hat{\rho}_{t}^{\prime }}{\partial t}=
[(\hat{H}_{0}+ \widehat{\widetilde{J}}+ \hat{H}_{f}),
(\widehat{\widetilde{\rho }}+ \hat{\rho}_{t}^{\prime })]$ and its
solution for $\hat{\rho}_{t}$ is sought with an accuracy of the
first order in $\hat{H}_{f} \sim E \sim \kappa \lambda ^{2}$,
assuming that $\widehat{\phi }^{0} \sim \kappa $ and
$\widehat{\widetilde{J}} \sim \lambda$, where $\lambda$ is the
scattering parameter. In this  approximation, the transition to
the interaction representation [$\widehat{A}(t)=
e^{i\hat{H}_{0}t}\widehat{A}e^{-i\hat{H}_{0}t}$] provides
\begin{gather}
\rho_{t}^{\prime}= (-i) \left \{ \int_{0}^{t}[H_{f}(\tau -t),
\hat{\rho}^{0}] d\tau+ \int_{0}^{t}[\widehat{\widetilde{J}}(\tau
-t),\widehat{\widetilde{\rho }}] d\tau \right \} \nonumber \\
+(-i)^{2} \int_{0}^{t}d\tau \int_{0}^{\tau }d\tau ^{\prime
}[\widehat{\widetilde{J}}(\tau -t),[\widehat{\widetilde{J}}(\tau
^{\prime }-t),\widehat{\widetilde{\rho }}]]. \label{ro3}
\end{gather}

The conditions of the quasistationarity  of the current density
matric $\widehat{\widetilde{\rho}}$ in the limit of infinite time
reduce to $\overline{\hat{F}_{l}^{s}}|_{t\rightarrow \infty}=
\overline{\hat{F}_{l}^{s}}|_{t=0}$, i.e., to the system of
equations
\begin{equation} \label{ro4a}
Sp\{\rho _{t}^{\prime }\hat{F}_{l}^{s}\}|_{t\rightarrow \infty}=0.
\end{equation}

In the limit of the infinite number of moments, system
(\ref{ro4a}) is equivalent to the exact kinetic equation for the
nonequilibrium one- particle density matrix. The equations of
system (\ref{ro4a}) have the usual kinetic form
\begin{gather}
\left \{ i\int_{0}^{t}d\tau Sp\{\hat{F}_{l}^{s}[H_{f}(\tau
-t),\hat{\rho}^{0}] \right. \nonumber \\
\left . + \int_{0}^{t}d\tau \int_{0}^{\tau }d\tau ^{\prime
}Sp\{\hat{F}_{l}^{s}[\widehat{\widetilde{J}}(\tau
-t),[\widehat{\widetilde{J}}(\tau ^{\prime }-t),\widehat{\phi
}^{0}]] \right\}_{t\rightarrow \infty }=0, \label{ro4b}
\end{gather}
where the first and second terms correspond to the field and
collision terms, respectively. We denote these terms as
$tX_{l}^{s}$ and $tP_{l}^{s}$, respectively. Equations
(\ref{ro4a}) determine coefficients $\eta _{l}^{s}$ in
$\widehat{\phi }^{0}$.

The detailed form of the collision term in Eq.(\ref{ro4b})
includes  expression of the form $Sp\{ \hat{F}_{l}^{s}
\Lambda_{\mathbf{q}_{1}}^{\alpha}(\widetilde{\tau})
S_{\mathbf{q}_{1}}^{\alpha}(\widetilde{\tau })
\Lambda_{\mathbf{q}_{2}}^{\beta}(\widetilde{\tau }^{\prime})
S_{\mathbf{q}_{2}}^{\beta }(\widetilde{\tau }^{\prime})
\hat{\rho}^0 \hat{F}_{l_1}^{s_1} \}$, where $\widetilde{\tau}=
\tau -t$ and $\widetilde{\tau }^{\prime }= \tau -t$, which are
calculated with the use of the mode- coupling approximation
\cite{Plakida97}. In this approximation, "outer"
$S_{\mathbf{q}}^{\alpha }$ operators [i.e.,
$S_{\mathbf{q}}^{\alpha }$ operators entering into
$\widehat{\widetilde{J}}$ in Eqs. (\ref{5})] are separately
averaged at the first stage and the initial averaging is performed
for them. The unaveraged averages of the remaining operators that
appear after the above procedure are calculated at the next stage:
$Sp {\,} \{ \hat{F}_{l}^{s} \Lambda_{\mathbf{q}_{1}}^{\alpha
}(\widetilde{\tau }) S_{\mathbf{q}_{1}}^{\alpha }(\widetilde{\tau
}) \Lambda _{\mathbf{q}_{2}}^{\beta }(\widetilde{\tau }^{\prime})
S_{\mathbf{q}_{2}}^{\beta }(\widetilde{\tau }^{\prime})
\hat{\rho}^0 \hat{F}_{l_1}^{s_1} \}$ $=>$ $\left( Sp {\,}
\hat{F}_{l}^{s}
\Lambda_{\mathbf{q}_{1}}^{\alpha}(\widetilde{\tau})
\Lambda_{\mathbf{q}_{2}}^{\beta }(\widetilde{\tau }^{\prime})
\hat{\rho}^0 \hat{F}_{l_1}^{s_1} \right) \left( Sp {\,}
\hat{\rho}^{0} S_{\mathbf{q}_{1}}^{\alpha }(\widetilde{\tau})
S_{\mathbf{q}_{2}}^{\beta }(\widetilde{\tau }^{\prime }) \right)$

As a result, for the collision term, we obtain
\begin{equation}
P_{l}^{s}=2N {\,} \sum\limits_{l_{1}}\eta
_{l_{1}}^{s_{1}}P_{ll_{1}}^{s,s_{1}}, \label{P1}
\end{equation}

\begin{gather}
P_{ll_{1}}^{s,s_{1}}=J^{2}\frac{1}{N^{2}} {\,} \sum\limits_{\atop
{\mathbf{k},\mathbf{q}}{s,s^{\prime}}} \left(
F_{l}^{s}(\mathbf{k})- F_{l}^{s}(\mathbf{k+q})\right) \nonumber
\\
\times \left(F_{l_{1}}^{s_{1}}(\mathbf{k})-
F_{l_{1}}^{s_{1}}(\mathbf{k+q})\right) Z_{\mathbf{k}}^{(s)}
Z_{\mathbf{k+q}}^{(s_{1})} \nonumber \\
\times
n_{F}(E_{\mathbf{k}}^{(s)})[1-n_{F}(E_{\mathbf{k+q}}^{(s_{1})})]
\nonumber \\
\times n_{B}(E_{\mathbf{k+q}}^{(s)}- E_{\mathbf{k}}^{(s_{1})})
\chi^{\prime \prime} (\mathbf{q},(E_{\mathbf{k+q}}^{(s_{1})}-
E_{\mathbf{k}}^{(s)})) \label{P2}
\end{gather}
where $\chi^{\prime \prime}(\mathbf{q},\omega )$ is the imaginary
part of spin susceptibility and $n_{B}(\omega )$ is the Bose
function.

Under the assumption that $\hat{F}_{l}^{s}$ in the expression for
$\widehat{\phi }^{0}$ in Eq. (\ref{ro2}) is quasidiagonal, because
$H_f$ in Eqs. (\ref{6}) is quasidiagonal, the field term
$X_{l}^{s}$ in Eq. (\ref{ro4b}) is modified to the form
\begin{equation}
X_{l}^{s}=E\sum_{\mathbf{k}}V_{\mathbf{k}}^{x(s)}\left(
\frac{-\partial n_{F}(E_{\mathbf{k}}^{(s)})}{\partial
E_{\mathbf{k}}^{(1)}}\right)
Z_{\mathbf{k}}^{(s)}F_{l}^{s}(\mathbf{k)}. \label{X1}
\end{equation}

Finally, the average current is obtained in the form
\begin{gather}
j^{x}= \sum_{l,s} {\,} \eta_{l}^{s} Sp{\,} \{
\hat{\rho}\hat{F}_{l}^{s}[\hat{H}_{0}, \hat{x}] \}=
\frac{2}{g^{2}a_{z}N} \nonumber \\
\times \sum_{l,s,\mathbf{k}} \eta_{l}^{s} F_{l}^{s}(\mathbf{k)}
V_{\mathbf{k}}^{x(s)} Z_{\mathbf{k}}^{s} n_F(E_{\mathbf{k}}^{(s)})
[1- n_F(E_{\mathbf{k}}^{(s)}) ], \label{j1}
\end{gather}
where $a_{z}$ is the distance between $CuO_{2}$ planes (we take
$a_{z}=6.6$ A and the corresponding volume of the unit cell is
$g^{2}a_{z}=93$ A$^{3}$).

With the use of the coefficients $\eta _{l}^{s}$ obtained from
equation $X_{l}^{s}+ P_{l}^{s}=0$ (\ref{ro4b}), the current
density and diagonal component of the resistivity tensor $\rho $
can be determined.

For the doping case of interest, $n_{h}\lesssim 0.2$ the chemical
potential $\mu$ lies sufficiently deep in the lower polaron band.
For this reason, only the lower band $s=1$ can be retained in the
expressions for the current and field and collision term; thus,
the summation over $s$ is removed in Eqs. (\ref{P2} and (\ref{j1}.
Thus, the problem reduces to a usual one- band case, where the
spectrum of the lower polaron band provides the characteristics of
the carrier spectrum. The significant difference from traditional
expressions is that the residue function $Z_{\mathbf{k}}^{(1)}$
explicitly enters into kinetic equations (\ref{P2}) -- (\ref{j1})
and into Eq. (\ref{a5}) for $n_{h}$. Only the lower polaron band
will be taken into account and its index 1 will be omitted in the
notation of energy, residues, velocities, and coefficients $\eta$
and moments $F_{l}(\mathbf{k)}$ in Eqs. (\ref{ro2}).

It is known that the dynamics of charge carriers in $CuO_{2}$
planes is well described by the three- band Emery model
\cite{Emery87_88, Zhang88}. The calculation of the spin polaron
spectrum with the use of the Emery model \cite{Bar2001} provides
the spectrum observed in the ARPES experiments in a wide doping
range. The assumption of the correspondence between doping in
models with free carriers and frustration $p$ in the pure spin
model \cite{Inui88} is a key assumption for the description of the
properties of the lower polaron band. This assumption is
physically natural: a moving hole destroys the magnetic order and
the same effect occurs with increasing $p$ in the pure spin model.
Moreover, it is based on the similar character of changes in spin
correlation functions as functions of $x$ and $p$. However, strict
statements on the $x\leftrightarrow p$ correspondence are absent,
but frustrations is always present in the spin subsystem at the
doped $CuO_{2}$ plane. Even in the dielectric limit, the ratio of
exchange  on the second neighbors to exchange on on the first
neighbors is estimated as $I_{2}/I_{1} \approx 0.1$
\cite{Annet89}. The role of frustration as a driving force of the
formation of various spin- liquid states is widely discussed. It
is expected that a quantum phase transition can occur near
$I_{2}/I_{1}\approx 0.5$ (which corresponds to $p\approx 0.3$)
\cite{LeeRMP06}. Close frustration parameter values are accepted
when discussing the stripe scenario of the appearance of
incommensurate peaks \cite{Sachdev04+}.

A decrease in the spin correlation length $\zeta $ with increasing
$p$ corresponds to change in the spin correlation functions, which
explicitly appear in the equations for the Green's functions of
the spin polaron and significantly affect the spectrum
$E_{\mathbf{k}}$ and residue function $Z_{\mathbf{k}}$. Below, we
will discuss the problem of the electrical resistivity for the
cases $A$ and $B$ of a small spin correlation length on the order
of several lattice constants $\zeta \simeq g$ ($p\simeq 0.3$) and
a large spin correlation length $\zeta \gtrsim 10g$ ($p\simeq
0.1$), respectively. Cases $A$  and $B$ refer to cuprates close to
optimal doping and to strongly undoped cuprates.

\begin{figure} \label{fig1}
\includegraphics[width=8cm]{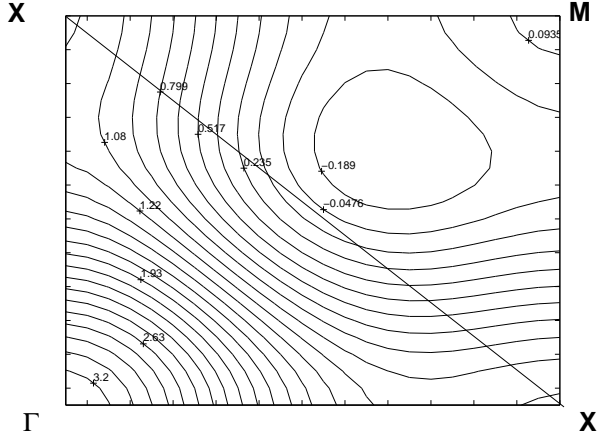}
\caption {Fig. 1. Hole spectrum $E_{\mathbf{k}}^{A}$ (\ref{E_k})
presented by contours $E_{\mathbf{k}}^{A}= const$ in units $\tau
=0.2$ $eV$ in the first quarter of the Brillouin zone.}
\end{figure}

\begin{figure} \label{fig2}
\includegraphics[width=8cm]{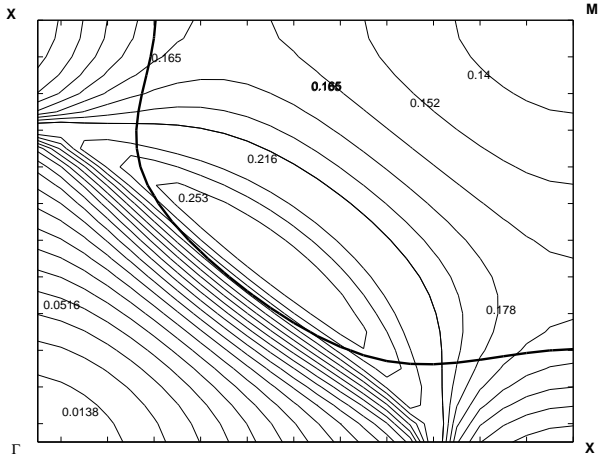}
\caption {Fig. 2. Residue function $Z_{\mathbf{k}}^{A}$ (doping is
close to optimal) presented by contour curves in the first quarter
of the Brillouin zone. The thick line is the Fermi surface in case
A.}
\end{figure}

\begin{figure} \label{fig3}
\includegraphics[width=8cm]{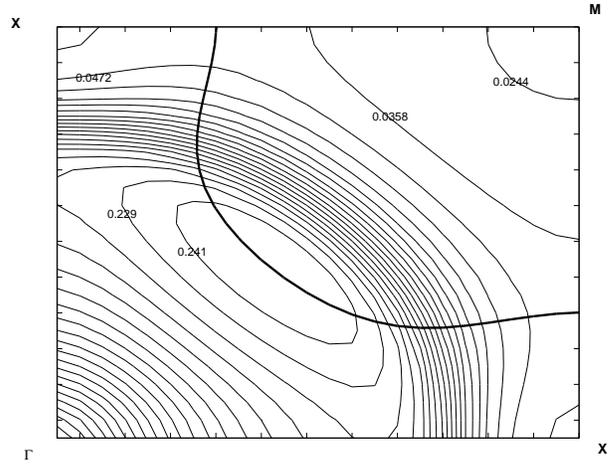}
\caption {Fig. 3. Residue function $Z_{\mathbf{k}}^{B}$ (low
doping) presented by contour curves in the first quarter of the
Brillouin zone. The thick line is the Fermi surface in case B.}
\end{figure}

Spin fermion Hamiltonian (\ref{1}) applied to the Emery model for
case $A$ provides the spectrum $E_{\mathbf{k}}^{A}$ and residue
function $Z_{\mathbf{k}}^{A}$ \cite{Bar2001}, whose characteristic
form is shown in Figs. 1 and 2, respectively. The analytical form
of $E_{\mathbf{k}}^{A}$ is approximated with the use of the
square- symmetry harmonics
\begin{gather}
E_{\mathbf{k}}^{A}= \tau (a_{1}\gamma _{g}(\mathbf{k})+
a_{2}\gamma _{g}^{2}(\mathbf{k}) \nonumber \\
+ a_{3}\gamma _{d}(\mathbf{k})+ a_{4}\gamma_{d}^{2}(\mathbf{k})+
a_{5}\gamma _{g}(\mathbf{k})\gamma _{d}(\mathbf{k})); \label{E_k}
\end{gather}
where
\begin{gather}
\gamma _{g}(\mathbf{k})= (\cos gk_{x}+\cos gk_{y})/2, \\
\gamma _{d}(\mathbf{k})=\cos gk_{x}\cos gk_{y},
\end{gather}
and $a_{1}=1.5$, $a_{2}=3.0$, $a_{3}=-1.25$, $a_{4}=0.0$, and
$a_{5}=0.1$. The following energy- parameter values characteristic
of the Emery model are assumed below: $\tau =0.2$ $eV$, $J=\tau$,
$I=0.5\tau$, and temperature $T=0.1I\approx 120$ $K$.

The form of $Z_{\mathbf{k}}^{B}$ is shown in Fig. 3. The form of
the spectrum $E_{\mathbf{k}}^{B}$ qualitatively corresponds to the
following scenario of the evolution of $E_{\mathbf{k}}$ with
increasing $\zeta $: the bottom of the spectrum
$E_{\mathbf{k}}^{A}$ (see Fig. 1) is shifted along the
$\mathbf{\Gamma }\leftrightarrow \mathbf{M}$ diagonal toward the
point $\mathbf{\Gamma }(0,0)$. This shift at low doping ensures a
"large" Fermi surface located near the magnetic Brillouin zone
$\mathbf{X}\leftrightarrow \mathbf{X}$. This scenario is confirmed
both by ARPES experiments and by calculations of the spin polaron
spectrum. In this case, the spectrum $E_{\mathbf{k}}^{B}$ near the
Fermi surface and magnetic Brillouin zone is inevitably more flat
than the spectrum $E_{\mathbf{k}}^{A}$. This flattering can be
treated as the narrowing of the band with increasing the polaron
effect (decreasing frustration). In order to taken into account
this narrowing, we set $E_{\mathbf{k}}^{B}= 0.75 {\,}
E_{\mathbf{k}}^{A}$. The electrical resistivity is calculated for
the Fermi surfaces shown by thick lines in Figs. 2 and 3 for case
$A$ ($\mu^A= 0.82 \tau$) and $B$ ($\mu^B= 0.375 \tau$),
respectively.

In order to explicitly clarify the role of the flattering of
$E_{\mathbf{k}}$ and the $\mathbf{k}$ dependence of
$Z_{\mathbf{k}}$, we use the following spin susceptibility for
both cases
\begin{equation}
\chi (\mathbf{q},\omega )= \frac{-A_{\mathbf{q}}}{\omega
^{2}-\omega _{\mathbf{q}}^{2}+i\omega \gamma}.
\end{equation}
This form of the spin susceptibility appears in the framework of
the spherically symmetric self- consistent approach
\cite{Shimahara91, BarBerez94-both} in the method of irreducible
two- time retarded Green's functions $\langle \langle
S_{\mathbf{q}}^{\alpha }|S_{\mathbf{-q}}^{\alpha }\rangle \rangle
_{\omega +i\epsilon }$ \cite{Tserkovnikov71, Plakida73},
\cite{BarBerez94-both} or in the memory function method
\cite{Prelovsek04}. The spin susceptibility is taken in the same
form as in the case of the calculation of the electrical
resistivity at $T=0.1I$ in \cite{Bar2007}, where the method of
self- consistent calculation of $\chi (\mathbf{q},\omega )$ and
choice of damping $\gamma =0.45I$ were discussed in detail.

To take into account strong scattering anisotropy due to the
strong scattering of carriers from the spin mode with the
antiferromagnetic vector $\mathbf{Q}$, it is inevitably necessary
to go beyond the framework of the traditional one moment
approximation when solving the kinetic equation. The large number
of moments is also necessary for demonstration of the convergence
of the method.

Polynomials of the velocity components $V\mathbf{_{k}^{\alpha }}$
and their derivatives will be used below as the moments
$F_{l}(\mathbf{k})$ of the distribution function:
\begin{gather}
F_{l}^{E}(\mathbf{k})=\{V_{\mathbf{k}}^{x},{\,}(V_{\mathbf{k}}^{y})^{2}V_{\mathbf{k}}^{x},{\,}\frac{\partial
V_{\mathbf{k}}^{x}}{\partial k^{y}}V_{\mathbf{k}}^{y},{\,}
\nonumber \\
\frac{\partial V_{\mathbf{k}}^{y}}{\partial
k^{y}}V_{\mathbf{k}}^{x},{\,}\frac{\partial
V_{\mathbf{k}}^{x}}{\partial k^{x}}\frac{\partial
V_{\mathbf{k}}^{y}}{\partial
k^{y}}V_{\mathbf{k}}^{x},{\,}(V_{\mathbf{k}}^{x})^{3}{\,}
\frac{\partial V_{\mathbf{k}}^{x}}{\partial
k^{x}}V_{\mathbf{k}}^{x}\}. \label{momX}
\end{gather}

For good convergence, it is sufficient to take into account first
three or four moments.

Calculations for case $A$ give the electrical resistivity
$\rho_{A}=$ $85.2$ $\mu \Omega cm$ and the number of holes $n_{h}=
0.21$. This value is comparable with $\rho (T=120K)\approx $ $100$
$\mu \Omega cm$ for $La_{2-x}Sr_{x}CuO_{4}$ with $0.16 < x < 0.22$
\cite{Ando04}. The electrical resistivity $\rho_A$ calculated
under the assumption $Z_{\mathbf{k}}^{A}= 1$ is denoted as
$\widetilde{\rho}_{A}$ (below, all the quantities calculated under
the assumption $Z_{\mathbf{k}}^{A}=1$ are marked by tilde) and is
$\widetilde{\rho }_{A}= 84.2\mu \Omega cm \approx \rho _{A}$ and
$\widetilde{n}_{h}\approx 1.13$. The values $\widetilde{\rho}_A$
and $\rho _{A}$ are approximately equal to each other, because
$Z_{\mathbf{k}}^{A}$ depends only slightly on $\mathbf{k}$ in the
$\mathbf{k}$- space Brillouin- zone region that it is near the
Fermi surface and contributes to kinetics. In this region,
$Z_{\mathbf{k}}\approx $ $const= Z$. In this case, $\eta _{l}\sim
Z^{-1}$ according Eqs. (\ref{P1})- (\ref{X1}) and the resistivity
$\rho$ determined by the current (\ref{j1}) is independent of $Z$.
In this case, the results obtained in the spin- fermion models,
where the residue function is disregarded, are valid. However, the
smallness of $Z_{\mathbf{k}}^{A}$ ($Z_{\mathbf{k}}^{A}$ $\lesssim
0.2$), as well as the $\mathbf{k}$ dependence of $Z_{\mathbf{k}}$
in the entire $\mathbf{k}$- space Brillouin zone, see Fig. 2,
obviously leads to the relation $n_{h}\lesssim 0.2$ at a large
Fermi surface and affects the position of the Fermi surface with
respect to the magnetic Brillouin zone. The last effect
significantly determines the collision integral for strong
scattering by the antiferromagnetic vector $\mathbf{Q}$.

In case $B$, $Z_{\mathbf{k}}^{B}$ and the Fermi surface shown in
Fig. 3 correctly represent the properties of the spectrum of
underdoped cuprates: the hole residues $Z_{\mathbf{k}}^{B}$
decreases from 0.24 to 0.04 when moving along the Fermi surface
from the point of intersection with the $\mathbf{Ã-M}$ to thre
point of intersection of the Fermi surface with the  boundary
$\mathbf{X-M}$ of the Brillouin. This decrease qualitatively
reflects the known opening of the pseudogap on the Fermi surface.
Calculations for the case $B$ lead to $\rho _{B}=$ $231.1$ $\mu
\Omega cm$ and the number of holes $n_{h}=0.08$, which is close to
$\rho (T= 120K) \approx $ $220$ $\mu \Omega cm$ for
$La_{2-x}Sr_{x}CuO_{4}$ with $x\approx 0.1$ \cite{Ando04}. With
$Z_{\mathbf{k}}=1$, we obtain $\widetilde{\rho }_{B}= 162.1\mu
\Omega cm \approx \rho_A$ and $\widetilde{n}_{h}\approx 0.92$.
Thus, the inclusion of the $\mathbf{k}$ dependence of
$Z_{\mathbf{k}}$ becomes important in the case of low doping.
Moreover, the empirical approximation $n_{h}\approx
1-\widetilde{n}_{h}$ is completely violated.

Our calculations also demonstrate the strong dependence of $\rho $
on the band narrowing, which is described by the Drude formula
$\rho= m^*/ne^2\tau$ for simple metals. Indeed, the electrical
resistivity calculated with the unflattened spectrum
$E_{\mathbf{k}}^{A}$ (the rigid band, but with residues
$Z_{\mathbf{k}}^{B}$, see Fig. 3) for the Fermi surface position
in the Brillouin zone shown in Fig. 3 is equal to
$\rho_B^{\prime}=$ $124.9$ $\mu \Omega cm$ ($\ll 231.1$ $\mu
\Omega cm$); in this case, $n_{h}$ remains unchanged:
$n_{h}=0.08$. The calculation for the case $B$ (we recall that the
Fermi surface in case $B$ is always located as in Fig. 3) and the
rigid band $E_{\mathbf{k}}= E_{\mathbf{k}}^{A}$ in the "rigid"
residue approximation $Z_{\mathbf{k}}=$ $Z_{\mathbf{k}}^{A}$ gives
lower electrical resistivity $\rho _{B}^{\prime \prime }=$ $89.5$
$\mu \Omega cm$. The calculation for the rigid band and
$Z_{\mathbf{k}}= 1$ provides $\widetilde{\rho }_{B}^{\prime}=$
82.2.5 $\mu \Omega cm$ and $\widetilde{n}_{h}\approx 0.92$.

Thus, we not only have derived the kinetic equation for the spin-
polaron carriers, but also have shown that both change in the
residue function $Z_{\mathbf{k}}$ and band narrowing must be
included in the description of kinetics on the basis of the Fermi
surface obtained from the ARPES measurements for various doping
degrees (e.g., the Fermi surface shown in Figs. 2 and 3).

\end{document}